\documentclass[12pt]{article}
\usepackage{epsfig,pslatex}

\begin{document}

\baselineskip 20pt

\begin{flushright}
{\bf SDU-HEP200701} \\
%{\bf hep-ph/0701212}
\end{flushright}

\vskip 20pt

\begin{center}

{\LARGE{\bf Doubly Heavy Baryon Production at $\gamma \gamma$ Collider}}\\
\vspace{20pt}
{\small \it Shi-Yuan Li\footnote{lishy@sdu.edu.cn},
Zong-Guo Si\footnote{zgsi@sdu.edu.cn},
Zhong-Juan Yang\footnote{yangzhongjuan@mail.sdu.edu.cn}\\
Department of Physics, Shandong University, Jinan, 250100,
P. R. China}

\vspace{20pt}
{\bf Abstract}
\end{center}
\noindent
The inclusive production of doubly heavy baryons $\Xi_{cc}$ and
$\Xi_{bb}$ at $\gamma\gamma$
collider is investigated. It is found that the
contribution from the  heavy quark pair $QQ$ in color triplet and
color sextet are important. %\\[10pt]


%\vspace{0.2cm}

\section{Introduction}
%\vspace{0.3cm}

 $\gamma\gamma$ collider is one option of the International
Linear Collider (ILC) in the future, where many interesting issues
can be employed\cite{issues}.
The doubly heavy baryon production is a potential one. It can
be factorized into two stages, i.e., the production of two heavy quarks
and their transformation into the baryon.
The heavy quark mass sets the large scale, which enables us to  calculate
their creation within Perturbative QCD framework. The transformation is
nonperturbative, and may be dealt with non-relativistic QCD (NRQCD)
because of the small velocity of the heavy quark
in the rest frame of the baryon \cite{Bodwin:1994jh}.
This  provides us a good opportunity to study QCD, both in the
perturbative and nonperturbative aspects.
For this aim, a fully inter-course between phenomenological
and experimental investigations in various processes is required.

The doubly heavy baryon $\Xi_{cc}$ has been observed by
SELEX collaboration
\cite{Mattson:2002vu,Moinester:2002uw,Ocherashvili:2004hi},
 and many theoretical studies have been done, e.g., in
refs.
\cite{Baranov:1995rc, Berezhnoi:1998aa, Ma:2003zk, Chang:2006eu, Chang:2006xp}.
However, up to now, we can not understand its
production mechanism sufficiently yet. As pointed out
in ref.\cite{Kiselev:2002an},
the observed production rate is much larger than most
of the theoretical predictions, which means that further investigation
on the production mechanism as well as
exploration for more experimental opportunities are still necessary.

For two identical heavy quark system, if one neglects the
relative orbital angular momentum, there are only two states due to
the anti-symmetry of the total wave function.  One
is in angular momentum $^3S_1$ and color triplet, and
the other is $^1S_0$ and color sextet. In general, both of
these two states can contribute to the doubly heavy baryon production,
which can be described by introducing two hadronic matrix elements
\cite{Ma:2003zk,Brambilla:2005yk}.
%In general, the baryon with two identical heavy quarks
%can be transformed from two states of the heavy quark pair, one
%is in angular momentum $^3S_1$ and color triplet,
%the other is $^1S_0$ and color sextet due to the anti-symmetry of the
%total wave function if one neglects the orbital angular momentum. This
%kind of transition can
%be described by introducing two hadronic matrix elements.
In refs.\cite{Baranov:1995rc, Berezhnoi:1998aa} the inclusive
production of  doubly heavy baryons  at
various colliders is studied, where only the contribution from quark pair
in $^3S_1$ and color triplet is included.
While in refs.\cite{Ma:2003zk,Chang:2006eu, Chang:2006xp}, the
contribution from both color triplet and color sextet are taken into account
for $e^+e^-$ and hadron-hadron colliders.
In this paper, our aim is to investigate the inclusive production of
doubly heavy baryons at the future $\gamma\gamma$ collider.

This paper is organized as follows: In Sec.\ref{sec2},
we list the basic formula
for $\gamma \gamma\to H_{QQ}+X$ and give some related numerical results.
The effective cross sections
for $\Xi_{cc}$ and $\Xi_{bb}$ production at ILC are presented
in Sec.\ref{sec3}.
Finally a short summary is given.

%\vspace{0.3cm}

\section{Doubly Heavy Baryon Production in $\gamma \gamma$ collisions}
\label{sec2}
%\vspace{0.3cm}

The doubly heavy baryon $H_{QQ}$ can be produced
via the following process
\begin{equation}
\label{wen1}
\gamma(p_1)+\gamma(p_2) \to H_{QQ}(k)+X ,
\end{equation}
where $p_1$, $p_2$ and $k$ respectively denote the four-momentum of the
corresponding particles. The unobserved state X can be divided into a
nonperturbative part $X_N$ and a perturbative part $X_P$. At tree level,
$X_P$ consists of two heavy anti-quarks $\bar{Q}\bar{Q}$.
 Adopting the notation in
ref.\cite{Ma:2003zk}, one can obtain the scattering amplitude for  $\gamma(p
_1)+\gamma(p_2) \to H_{QQ}(k)+\bar Q(p_3)+\bar Q(p_4)+X_N$  process
\begin{equation}
{\cal T} =\frac{1}{2}\int\frac{d^4k_1}{(2\pi)^4} A_{ij}(k_1,k_2,p_3,p_4)
\int d^4x_1 e^{-ik_1 \cdot x_1} \langle H_{QQ}(k)+ X_N| \bar{Q_i}(x_1)
\bar{Q_j}(0)|0 \rangle,
\end{equation}
where $k_1$, $k_2$ denote the  four-momentum of the
internal heavy quarks, and  $p_3$, $p_4$ the momentum of the anti-quarks.
i, j represent Dirac and color indices,
$Q(x)$ is the Dirac field for the heavy quark.

\begin{figure}
\begin{center}
\psfig{file=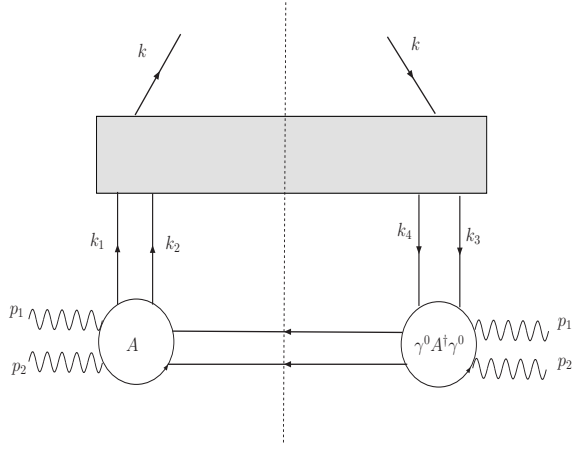, height=6cm, width=8cm}
\caption{Graphic representation for the contribution in eq.\ref{wen6}, the
dashed line is the cut and $k_4=k_1+k_2-k_3$,}
\label{fey4q}
\end{center}
\end{figure}

The differential cross section for $\gamma(p_1)+\gamma(p_2)
\to H_{QQ}(k)+\bar Q(p_3)+\bar Q(p_4)+X_N$  is
\begin{eqnarray}
\label{wen3}
&&d\hat{\sigma}(\hat s)=(\frac{1}{4})\frac{1}
{2\hat s} \sum_{X_N}
\frac{d^3 \bf k}{(2\pi)^3} \int \frac{d^3p_3}{(2\pi)^3 2E_3} \frac{d^3p_4}
{(2\pi)^3 2E_4}(2\pi)^4 \delta^4(p_1+p_2-k-p_3-p_4-P_{X_N}) \nonumber \\
&&\cdot \frac{1}{4} \int \frac{d^4k_1}{(2\pi)^4} \frac{d^4k_3}{(2\pi)^4}
A_{ij}(k_1,k_2,p_3,p_4) (\gamma^0 A^\dagger(k_3,k_4,p_3,p_4)\gamma^0)_{kl}
\nonumber \\
&& \cdot \int d^4x_1 d^4x_3 e^{-ik_1 \cdot x_1+ik_3 \cdot x_3} \langle
0|Q_k(0)Q_l(x_3)|H_{QQ}+X_N \rangle \langle H_{QQ}+X_N|\bar{Q_i}(x_1)
\bar{Q_j}(0)|0 \rangle, %\nonumber \\
\end{eqnarray}
where the average over polarization of the photons and the
summation over the spin of the baryon $H_{QQ}$ and over color, spin state
 of two $\bar Q$ quarks is implied, and $\hat s=(p_1+p_2)^2$.
The factor $1/4$ in the bracket is induced by
the identical photons and anti-quarks. The non-relativistic normalization for
the heavy baryon is used here. Employing the translation
invariance to eliminate the sum over $X_N$, one can  obtain
\begin{eqnarray}
\label {wen6}
d\hat{\sigma}(\hat s)&=&\frac{1}{2\hat{s}}\frac{d^3\bf k}{(2\pi)^3}
\int \frac{d^3p_3}
{(2\pi)^3 2E_3}\frac{d^3p_4}{(2\pi)^3 2E_4}\int \frac{d^4k_1}{(2\pi)^4}
\frac{d^4k_3}{(2\pi)^4} A_{ij}(k_1,k_2,p_3,p_4)\nonumber\\
&&\cdot (\gamma^0 A^\dagger(k_3,k_4,p_3,p_4)\gamma^0)_{kl}
\int d^4x_1 d^4x_2 d^4x_3 e^{{-ik_1}\cdot{x_1}-{ik_2}
\cdot{x_2}+{ik_3}\cdot {x_3}}\nonumber \\
&&\cdot \frac{1}{16} \langle0|Q_k(0)Q_l(x_3)a^\dagger({\bf k})
a({\bf k})\bar{Q_i}(x_1)\bar{Q_j}(x_2)|0\rangle,
\end{eqnarray}
where $a^\dagger({\bf k})$ is the creation operator for $H_{QQ}$ with
three momentum {\bf k}. This contribution can be represented by
Fig.\ref{fey4q}, where the black box represents the Fourier transformed
matrix element.
In the framework of NRQCD, at the zeroth order
of the relative velocity between heavy quarks in the rest frame
of $H_{QQ}$, the hadronic matrix element is \cite{Ma:2003zk}
\begin{eqnarray}
\label{wen9}
&&v^0 \int{d^4x_1 d^4x_2 d^4x_3 e^{{-ik_1}\cdot{x_1}-{ik_2}\cdot{x_2}
+{ik_3}\cdot {x_3}}\langle0|Q_k^{a_3}(0)Q_l^{a_4}(x_3)a^\dagger({\bf k})
a({\bf k})\bar{Q_i}^{a_1}(x_1)\bar{Q_j}^{a_2}(x_2)|0\rangle} \nonumber \\
&=&(2\pi)^4 \delta^4(k_1-m_Q v)(2\pi)^4
\delta^4(k_2-m_Qv)(2\pi)^4
\delta^4(k_3-m_Qv)\nonumber \\
&&\cdot [-{(\delta_{a_1a_4}\delta_{a_2a_3}+\delta_{a_1a_3}\delta_{a_2a_4})
(\tilde{P_v} C \gamma_5 P_v)_{ji}(P_v \gamma_5 C
\tilde{P_v})_{lk}}
\cdot{h_1} \nonumber \\
&&+(\delta_{a_1 a_4} \delta_{a_2 a_3} - \delta_{a_1 a_3} \delta_{a_2a_4})
(\tilde{P_v}C\gamma^{\mu} P_v)_{ji}(P_v \gamma^{\nu}
C\tilde{P_v})_{lk}
(v_{\mu}v_{\nu}-g_{\mu\nu})\cdot {h_3} ],
\end{eqnarray}
where $m_Q$ is the mass of the heavy quark, $a_i(i=1, 2, 3, 4)$ and
$i$, $j$, $k$, $l$ respectively denote the color and Dirac indices,
and
\begin{equation}
P_v=\frac{1+{\gamma}\cdot v}{2},\hspace{0.5cm}
C=i\gamma^2 \gamma^0,\hspace{0.5cm}
v^{\mu}=k^{\mu}/M_{H_{QQ}}.
\end{equation}
$\tilde {P_v}$ is the transpose of the matrix $P_v$.
$ h_1$ ($ h_3$) represents the probability
for a $QQ$ pair in  $^1S_0$ ($^3S_1$) state and in the color state of
$6$ ($\bar 3$) to transform into the baryon,
\begin{eqnarray}
h_1&=&\frac{1}{48}\langle0|[\psi^{a_1}\varepsilon \psi^{a_2}+\psi^{a_2}
\varepsilon \psi^{a_1}]a^{\dagger}a\psi^{{a_2}^{\dagger}}\varepsilon
\psi^{{a_1}^{\dagger}}|0\rangle,\nonumber\\
h_3&=&\frac{1}{72}\langle0|[\psi^{a_1}\varepsilon \sigma^n
\psi^{a_2}-\psi^{a_2}\varepsilon \sigma^n \psi^{a_1}]
a^{\dagger}a \psi^{{a_2}^{\dagger}}\sigma^n \varepsilon
\psi^{{a_1}^{\dagger}}|0\rangle,
\end{eqnarray}
where $\sigma^j(j=1, 2, 3)$ are Pauli matrices, $\varepsilon=
i\sigma^2$ is totally anti-symmetric, $\psi$ is the NRQCD field for
the heavy quark. Generally, $h_1$ and $h_3$ should be determined by
nonperturbative QCD.  Under NRQCD, $h_1$ and $h_3$ are at the same
order. $h_3$ can be related to the non-relativistic wave function at
the origin, i.e., $h_3=|\Psi_{QQ}(0)|^2$, which can be calculated in
the framework of potential models, such as in ref.
\cite{Bagan:1994dy}.

\begin{figure}
\begin{center}
\psfig{figure=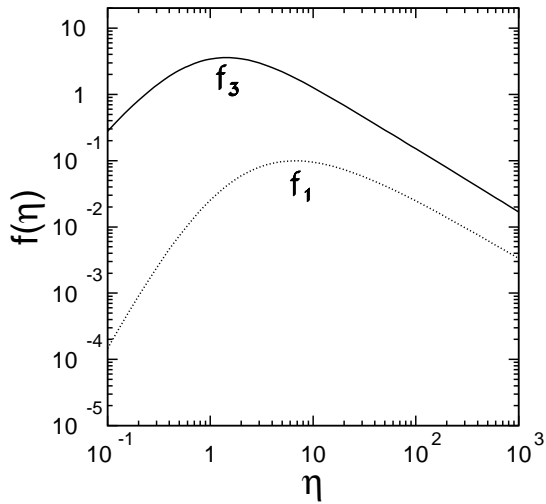,width=8cm}
\caption{The scaling function $f(\eta)$.}
\label{fhnpa}
\end{center}
\end{figure}

With all of the above equations, one can write the cross section for
$\gamma(p_1)+\gamma(p_2) \to H_{QQ}(k)+\bar Q(p_3)+\bar Q(p_4)+X_N$
process into the following form
\begin{equation}
\label{asde}
\hat{\sigma}=\frac{\alpha^2 \alpha_s^2 e_q^4}{4 m_Q^2} [f_1(\eta)
\frac {h_1}{m_Q^3}+ f_3(\eta) \frac {h_3}{m_Q^3}],
\end{equation}
where $\eta=\hat s/(16m_Q^2)-1$, $\alpha$ is
the fine structure constant, $\alpha_s$ the strong coupling, and
$e_q=2/3(-1/3)$ for up(down)-type quark.
The scaling functions $f_1(\eta)$ and $f_3(\eta)$ do not depend
on $m_Q$ explicitly. The numerical results for $f_1$ and $f_3$ are displayed
in fig.\ref {fhnpa}. One can find that the contribution from the color
triplet $Q  Q$ pair is much larger than that from
the color sextet one.

%\vspace{0.3cm}
\section{Effective cross sections at ILC}
\label{sec3}
%\vspace{0.3cm}

At a future linear collider, backscattered laser light may provide very
high-energy photons \cite{Ginzburg:1982yr}. The total effective cross
section at a photon collider is
\begin{equation}
\label{eeq14}
d\sigma (S)=\int_{0}^{y_{max}} dy_1 \int_{0}^{y_{max}} dy_2 f_{\gamma}^e
(y_1) f_{\gamma}^e(y_2) d\hat{\sigma}(\hat s),
\end{equation}
where $ \sqrt{S}$ is the $e^+e^- $ CMS energy. The function
$f_{\gamma}^e(y_1)$ is the normalized energy spectrum of the photons.
It is given by
\begin{equation}
\label{rre}
f_{\gamma}^e(y)={\cal N}^{-1}[\frac{1}{1-y}-y+(\frac{2y}{x (1-y)}-1)^2],
\end{equation}
where $\cal N$ is the normalization factor,
and $y$ is the fraction of the electron energy
transferred to the photon in the center-of-mass frame. It takes values in
the range
\begin{equation}
0\leq y \leq \frac{x}{x+1},
\end{equation}
where $x=4 E_L E_e/m_e^2$. $E_L$ ($E_e$) is the energy of the laser (electron)
beam and $m_e$ is the electron mass.  In order to avoid the creation of an
 $e^+e^-$ pair from the backscattered laser beam and the low energy laser beam,
the maximal value for $x$ is $2(1+\sqrt{2})$.
For a beam energy $E_e=250 GeV$, this leads to an optimal laser energy
$E_L= 1.26 eV$. Here, we adopt these two values for $E_e$, $E_L$.

\begin{figure}
\begin{center}
\psfig{figure=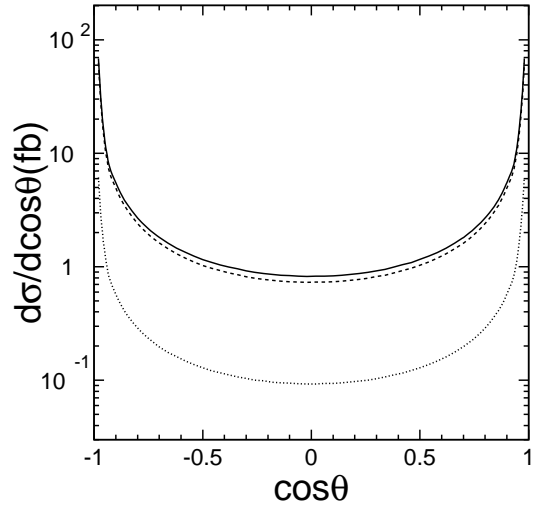,width=8cm}
\caption{ $cos\theta$-distributions, the solid line for
$h_3=h_1=|\Psi_{cc}(0)|^2$, the dotted for $h_1=|\Psi_{cc}(0)|^2$
and $h_3=0$, the dashed for $h_1=0$ and
$h_3=|\Psi_{cc}(0)|^2$.}
\label{efwtg}
\end{center}
\end{figure}

\begin{figure}
\begin{center}
\psfig{figure=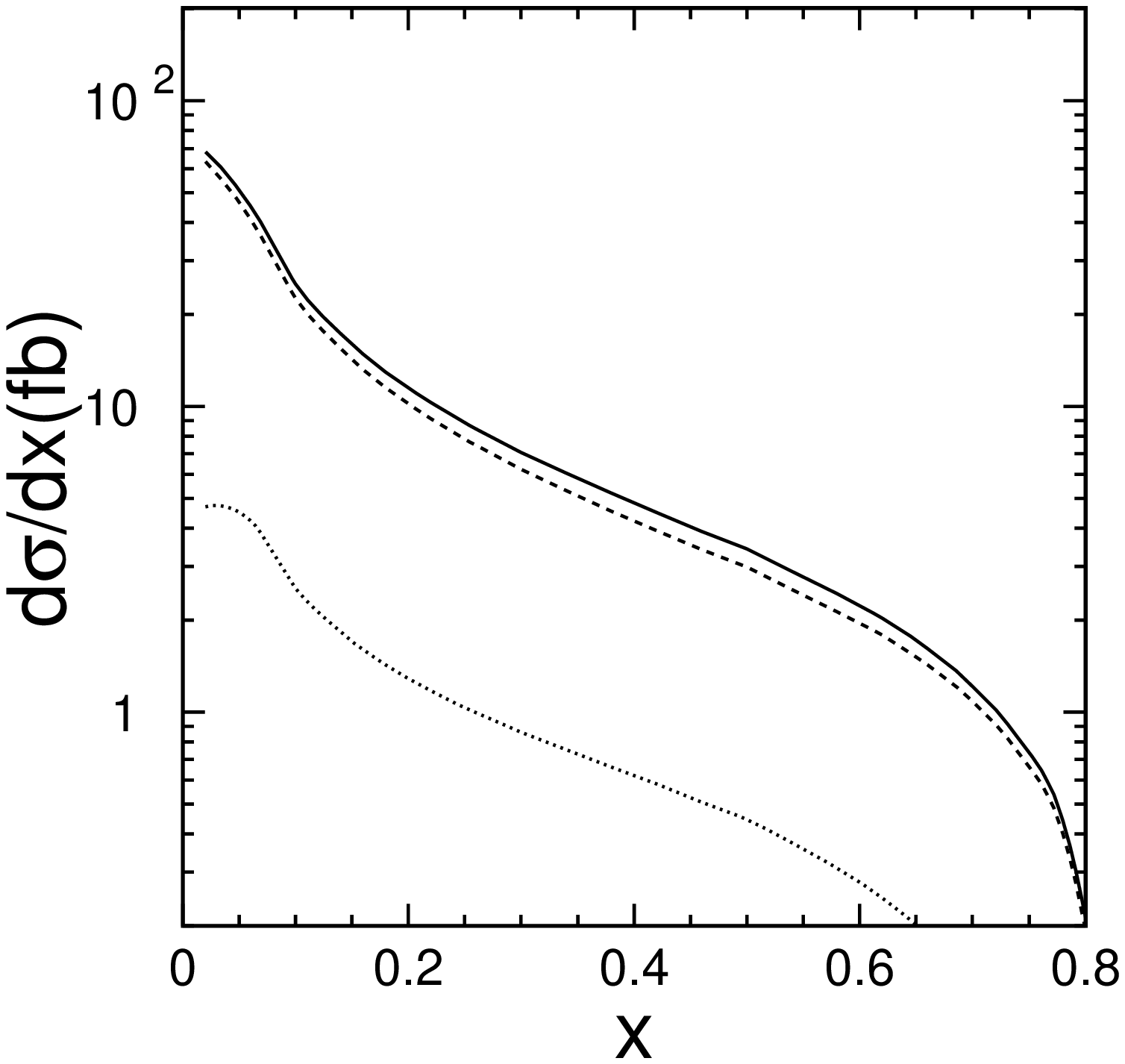,width=8cm}
\caption{Same as Fig.\ref{efwtg}, but for $x$-distributions.}
\label{errdj}
\end{center}
\end{figure}

\begin{figure}
\begin{center}
\psfig{figure=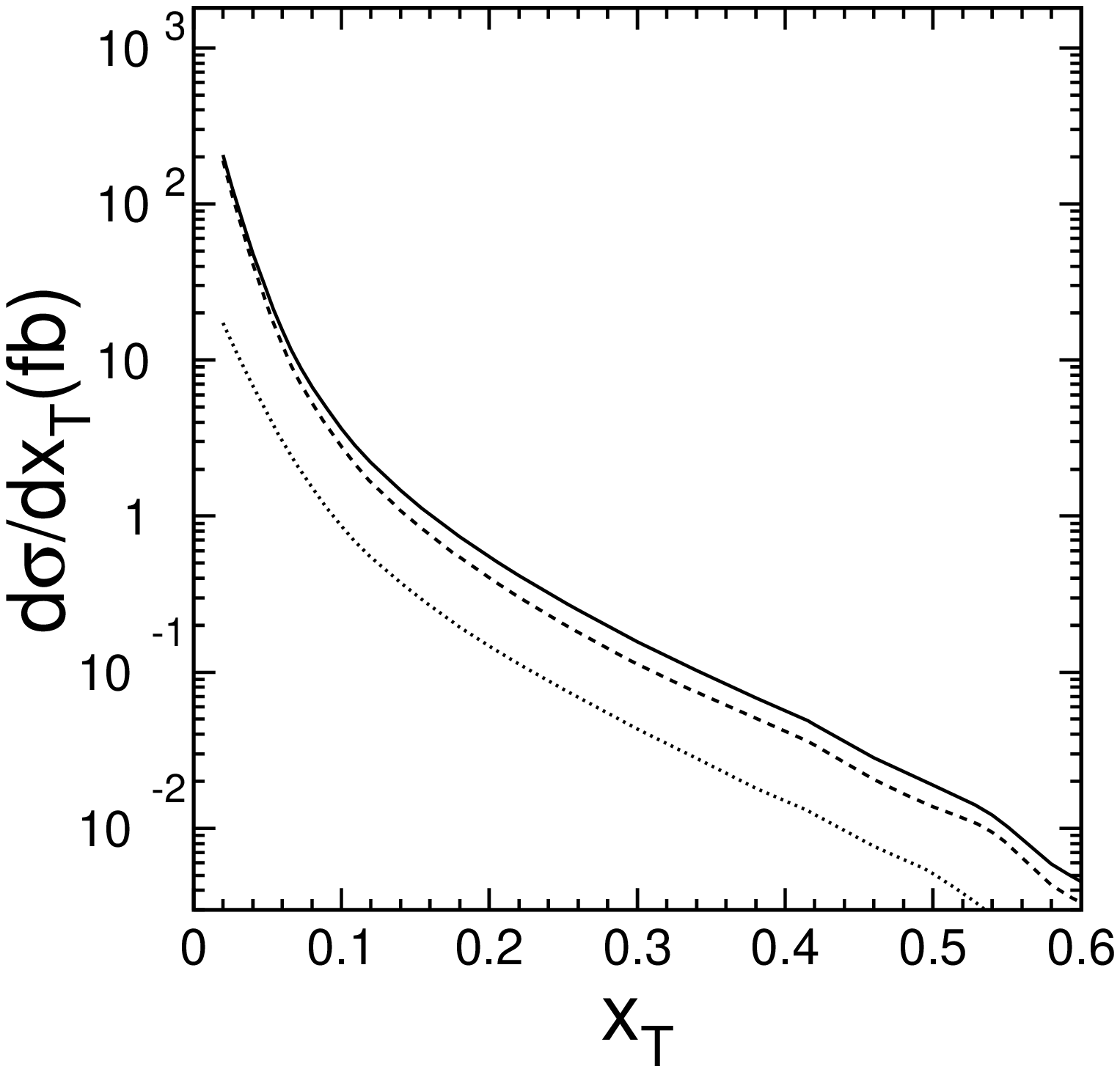,width=8cm}
\caption{Same as Fig.\ref{efwtg}, but for $x_T$-distributions.}
\label{xdew}
\end{center}
\end{figure}

%From eq. \ref{asde}, we can see that the cross sections depend on the value
%of the quark mass to certain inverse power. This can be checked by
%varying $m_Q$ and keeping $h_1$, $ h_3$ and $\alpha_s$ fixed. On the
%other hand, $\Psi_{QQ}(0)$ is also dependent on the value of quark mass.
In our numerical calculations, we take $|\Psi_{cc}(0)|^2=0.039$
 $GeV^3$ and $|\Psi_{bb}(0)|^2=0.152$ $GeV^3$ obtained in ref.
\cite{Bagan:1994dy}. For consistence, we adopt their value for quark
mass, i.e.,  $m_c=1.8 GeV$ and $m_b=5.1 GeV$. We choose the scale to
be $2 m_Q$ for $\alpha_s$, and obtain $\alpha_s=0.20(0.16)$ for
$\Xi_{cc}(\Xi_{bb})$ production. Our predictions for the total
effective cross section of $\Xi_{cc}$ $(\Xi_{bb})$ are given in
Table \ref{corr} (Table \ref{fdfde}). One can find that the
contribution from the color sextet $Q  Q$ pair are approximately $10
\%$ of that from the color triplet one if $h_1=h_3$. Obviously, the
effective cross section decreases as the heavy quark mass increases,
which is one reason for the production probability of $\Xi_{bb}$ is
much smaller than that of $\Xi_{cc}$. Additionally, taking
$\Xi_{cc}$ as an example, we give the predictions for the
distributions of $cos\theta$, $x$ and $x_T$. Here, $\theta$ and $x$
are defined in $e^+e^-$ CMS, where $\theta$ is the angle between the
moving direction of $\Xi_{cc}$ and that of the beam. $x=2E/
\sqrt{S}$ and $x_T=2P_T/ \sqrt{S}$, with $E$ and $P_T$ the energy
and transverse momentum of $\Xi_{cc}$ respectively. The
corresponding results are displayed in Fig.\ref{efwtg}, \ref {errdj}
and \ref {xdew} respectively.

%\begin{table}
%\begin{center}
%\begin{tabular}{|c|c||c|c|c|}  \hline
%  {\em $h_1(GeV^3)$} &{\em $h_3(GeV^3)$} & {\em $m_c=1.2(GeV)$}
%& {\em $m_c=1.5(GeV)$} & {\em $m_c=1.8(GeV)$}\\ \hline \hline
%$ 0$ & $0.039$  &139.19 (fb) & 63.42  (fb) & 33.21   (fb)\\ \hline
%$ 0.039$ & $0$    &  15.92 (fb)& 6.98 (fb)  & 3.54 (fb)\\ \hline
%$ 0.039$ & $0.039$ &  155.11(fb) & 70.40 (fb)  & 36.75(fb) \\ \hline
%\end{tabular}
%\caption{Results for the effective cross section of $\Xi_{cc}$ at
%$\sqrt {S}=500 GeV$.}
%\label{corr}
%\end{center}
%\end{table}

%\begin{table}
%\begin{center}
%\begin{tabular}{|c|c||c|c|}  \hline
%  {\em $h_1(GeV^3)$} &{\em $h_3(GeV^3)$} & {\em $m_b=4.8(GeV)$}
%& {\em $m_b=5.1(GeV)$} \\ \hline \hline
%$ 0$ & $0.152$  &0.227 (fb) &  0.181(fb) \\ \hline
%$ 0.152$ & $0$    &  0.0192 (fb)&  0.0150(fb) \\ \hline
%$ 0.152$ & $0.152$ & 0.2462  (fb)  &0.1960 (fb)   \\ \hline
%\end{tabular}
%\caption{Results for the effective cross section
%of $\Xi_{bb}$ at
%$\sqrt {S}=500 GeV$.}
%\label{fdfde}
%\end{center}
%\end{table}

\begin{table}
\begin{center}
\begin{tabular}{|c||c|c|c|}  \hline
  {\em $h_1(GeV^3)$} & $0$& $0.039$
&$0.039$ \\ \hline
 {\em $h_3(GeV^3)$} & $0.039$  &0 &0.039 \\ \hline \hline
 {\em $m_c=1.8(GeV)$}& 8.30 (fb) & 0.88 (fb)& 9.18 (fb)\\ \hline
\end{tabular}
\caption{Results for the effective cross section of $\Xi_{cc}$ at
$\sqrt {S}=500 GeV$.}
\label{corr}
\end{center}
\end{table}

\begin{table}
\begin{center}
\begin{tabular}{|c||c|c|c|}  \hline
  {\em $h_1(GeV^3)$} & $0$& $0.152$
&$0.152$ \\ \hline
 {\em $h_3(GeV^3)$} & $0.152$  &0 &0.152 \\ \hline \hline
 {\em $m_b=5.1(GeV)$}& 0.029 (fb) & 0.003 (fb)& 0.032 (fb)\\ \hline
\end{tabular}
\caption{Results for the effective cross section
of $\Xi_{bb}$ at
$\sqrt {S}=500 GeV$.}
\label{fdfde}
\end{center}
\end{table}

%\vspace{0.3cm}
\section{Summary}
%\vspace{0.3cm}

In this paper, we investigate the production of the doubly heavy
baryon $H_{QQ}$
at $\gamma \gamma$ collider. We find that for $H_{QQ}$ production, the
contribution from the color sextet is about 10\%. This implies that the
contribution from both color triplet and color sextet are important
for describing the doubly heavy baryon production.
Our method can be extended to
study the doubly heavy baryon $H_{QQ'}$ production. But for that case, more
hadronic matrix elements have to be considered due to no restriction from
Pauli principle for the heavy quark pair $QQ'$.
We hope our results can be helpful for the
investigation of the doubly heavy baryon production at the future linear
collider.

\section*{Acknowledgements}
This work is supported in part by NSFC, NCET of MoE and HuoYingDong
Foundation of China. The authors would like to thank Prof. J.P. Ma
(ITP) for his suggestions and helpful discussions, and also thank
all of the members in the theoretical particle physics group in
Shandong University for their helpful discussions. The author(Si)
would like to thank the hospitality of Phenomenology Group in
University of Wisconsin-Madison.

\end{document}